\documentclass[prd,a4paper,superscriptaddress,aps,showpacs,twocolumn,floats,nofootinbib]{revtex4}
\usepackage{epsfig}
\usepackage{graphicx}
\usepackage{amsmath}
\usepackage{amssymb}
\usepackage{color}
\usepackage{soul}
\usepackage{hyperref}
\usepackage{url}
\usepackage{float}

\newcommand{\be}{\begin{equation}}
\newcommand{\ee}{\end{equation}}

\newcommand{\bea}{\begin{eqnarray}}
\newcommand{\eea}{\end{eqnarray}}
\newcommand{\bean}{\begin{eqnarray*}}
\newcommand{\eean}{\end{eqnarray*}}

\begin{document}

\title{Birefringent Graphene Oxide Liquid Crystals in Micro-channels for Optical Switch}

\author{M. Arshadi Pirlar}
\email{m$_$arshadi@sbu.ac.ir}

\affiliation{Department of Physics, Shahid Beheshti University, Velenjak, Tehran 19839, Iran}
\affiliation{Ibn-Sina Multidisciplinary Laboratory, Department of Physics, Shahid Beheshti University, Velenjak, Tehran 19839, Iran}

\author{Y. Honarmand}

\affiliation{Department of Physics, Shahid Beheshti University, Velenjak, Tehran 19839, Iran}

\author{M. Rezaei Mirghaed}

\affiliation{Department of Physics, Shahid Beheshti University, Velenjak, Tehran 19839, Iran}

\author{S. M. S. Movahed}
\email{m.s.movahed@ipm.ir}

\affiliation{Department of Physics, Shahid Beheshti University, Velenjak, Tehran 19839, Iran}
\affiliation{Ibn-Sina Multidisciplinary Laboratory, Department of Physics, Shahid Beheshti University, Velenjak, Tehran 19839, Iran}
\author{R. Karimzadeh}
\email{r$_$karimzadeh@sbu.ac.ir}

\affiliation{Department of Physics, Shahid Beheshti University, Velenjak, Tehran 19839, Iran}
\affiliation{Ibn-Sina Multidisciplinary Laboratory, Department of Physics, Shahid Beheshti University, Velenjak, Tehran 19839, Iran}

\date{\today}

\begin{abstract}

We propose a mechanical-hydrodynamical experimental setup in which the microfluidic motion manipulates the optical birefringence of levitated graphene oxide liquid crystal. The birefringence of the sample is changed by flowing graphene oxide liquid crystal in the micro-channel. By measuring the ordinary and extraordinary refractive indices at five flow rates, one can determine the value of the birefringence of the samples, precisely. Our results demonstrate that, by adjusting the concentration and flow rate of dispersion of the graphene oxide nano flakes, the induced birefringence can be controlled. It is also shown that this approach can be used as an optical switch.
\end{abstract}


\maketitle

\vspace{0.05cm}
\section*{Introduction}

The control of arbitrary light polarization states is of major fundamental and technological relevance in photonics science and technology \cite{phillips2001,gao2003}. Polarization control is conventionally attained using anisotropic materials due to the birefringence. The anisotropic properties of materials could be manipulated by external stimuli such as applying electric or magnetic fields. One of the anisotropic materials are Liquid Crystals (LCs) which combine fluidity with birefringence, thereby enabling manipulation of the optical axis by external forces \cite{gennes1993}. Along with the recent enormous interest in graphene materials, graphene oxide-based (GO) LCs hold great promise for high-performance device operation. Indeed, dispersions of GO nano flakes in water exhibit liquid crystallinity in a specific range of GO concentrations \cite{xu2011}. It has been reported that the orientation of graphene oxide liquid crystal (GO-LC) can be changed by intervening a typical electric or magnetic fields \cite{shen2014,ahmad2015,ahmad2016,lin2017,ferrand2016,bao2017,changaei2019}. Adding the polymer component in the liquid solvent or by decorating the GO nano flakes with nanoparticles leads to retain the liquid crystallinity of GO.\\

Shen et al. illustrated that the direction of GO nano flakes directors is controlled by the electric field, utilized in the structure of a display using its Kerr effect  \cite{shen2014}. They also obtained the birefringence of GO-LC for different electric fields. Song et al. calculated the birefringence of GO-LC with the similar method but in a different experimental setup  \cite{ahmad2015}. They investigated the effect of solvents on the electro-optical switching and birefringence of GO disseminations \cite{ahmad2016}. Furthermore, the variation of the LC molecules orientational direction can be achieved with mechanical shear stress and fluid flow \cite{sengupta2013,sengupta2012}. With the advent of the microfluidic system, the orientation of the LC directors can be controlled by changing the geometry of the microfluidic channel \cite{sengupta2012,sengupta2012-1}. When the aspect ratio of channel is low, the orientation of the directors can be controlled due to the anchoring force and the cooperative alignment \cite{sengupta2012-1,sengupta2014,batista2015}. Actually, the interplay between steric interactions and the elastic energy of the director field is responsible for the anchoring behavior \cite{berreman1972}. The preferred orientation of director at a given interface is designated as easy axis \cite{batista2015}.\\

In our early study, we investigated whether the orientation and ordering of GO nano flakes can be manipulated by mechanical-hydrodynamical approach including fluid flow in micro-channel \cite{arshadi2019}. Surprisingly, it is observed that the transmitted and scattered intensity of light changed by controlling the flow rate due to the variation of the direction of flakes and birefringence. Therefore, our primary interest is to investigate the relation between the birefringence of the GO-LC and the fluid flow rate. For this purpose, we examined the effect of flow rate on the ordinary and extraordinary refractive indices (RI) of GO colloids. We also observed that the flow rate changed the RI and birefringence value of GO-LC. Investigations of the ordinary and extraordinary RI and birefringence of the dispersed GO nano flakes have been successfully carried out by total internal reflection (TIR) method. The birefringent materials were successfully used in a multitude of important optical devices \cite{ahmad2015,ahmad2016,ryu2017,kravets2015,konopsky2010,ye2013}. In this paper, the optical switching response of the sample is investigated.\\

\begin{figure}[H]
\centering
\includegraphics[width=0.3\textwidth]{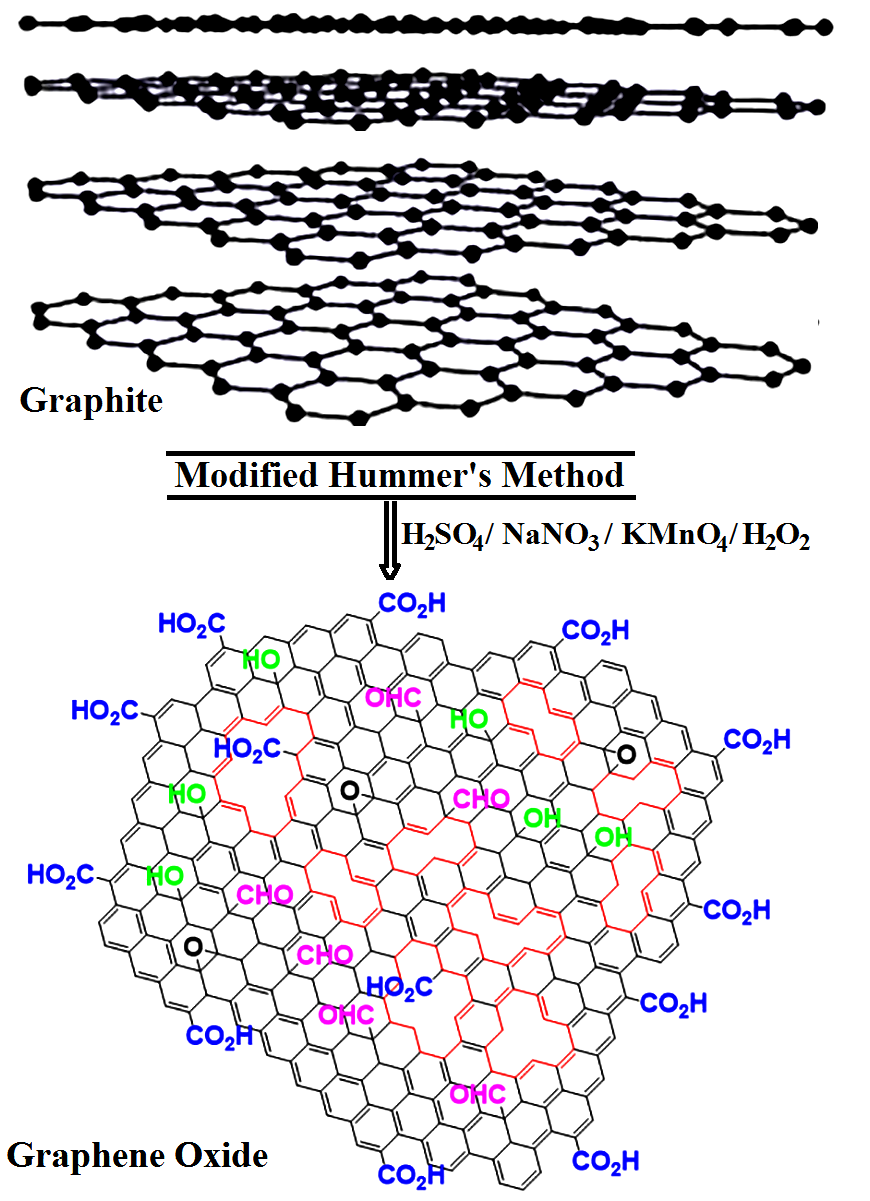}
     \caption{Schematic illustration of graphene oxide synthesis by chemical oxidation using modified Hummer's methods. }
\label{fig:schem}
\end{figure}

As demonstrated in Figure \ref{fig:schem}, GO nano sheets were synthesized through a modified hummer's method from graphite micro powder \cite{sepahvand2019}. The formation of graphene oxide dispersion is confirmed by dynamic light scattering (DLS), UV €"Visible (PerkinElmer LAMBDA 20), and Fourier transform infrared (FTIR, Perkin Elmer Spectrum Two) spectroscopes. 

\begin{figure}[H]
\centering
\includegraphics[width=0.5\textwidth]{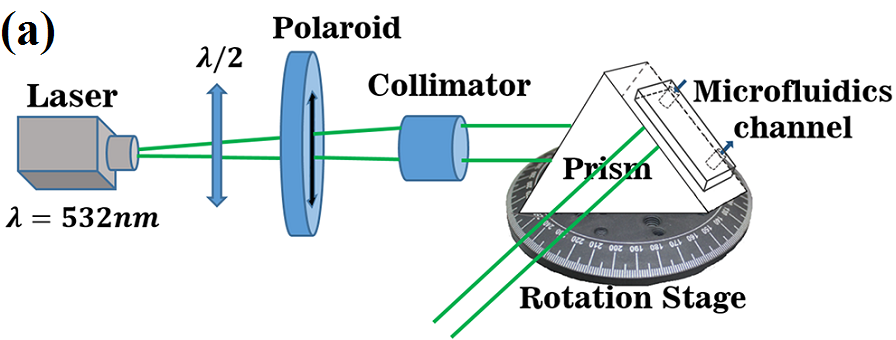}\\
\includegraphics[width=0.3\textwidth]{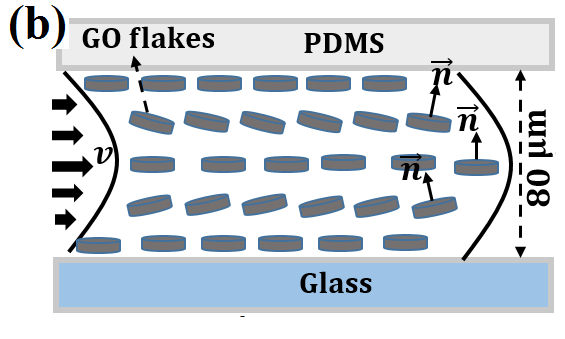}
     \caption{(a) A schematic for experimental setup, (b) The orientation of GO-LCs flakes in the channel walls ( $\vec{n}$ is the director of GO flakes).}
\label{fig:fig1}
\end{figure}

There are several methods for measuring the refractive index of materials such as total internal reflection (TIR), Ellipsometry, refractometry, surface plasmon resonance (SPR) and etc \cite{kravets2015,riviere1978,fujiwara2007,kravets2010}. Kravets et al. used the variable angle spectroscopic ellipsometry to measure the optical anisotropy of the graphene oxide films prepared by the microfluidic evaporation technique \cite{kravets2015}. In this technique, the microfluidic channel is used to create a micron sized film of GO nano sheets. They show that the fabricated films of the GO sheets have a large value of optical anisotropy in a wide spectral range. We used the TIR method that enable us to measure the ordinary and extraordinary RI of GO. In this technique, the critical reflection angle is used for measuring the ordinary and extraordinary RI of the sample \cite{konopsky2010}. The TIR was also used to calculate the absorption coefficient of a sandwiched graphene structure between a prism and a Polydimethylsiloxane (PDMS) layer \cite{ye2013}. Cheon et al. used the SPR and detection of the reflectance ratio at the critical TIR angle techniques to determine the complex RI of graphene \cite{cheon2014}. Also, the colorectal tissue complex indices were determined using a critical TIR angle approach \cite{giannios2017}. Accordingly, we demonstrate successful implementation of a reliable and robust approach to calculate the ordinary and extraordinary RI of levitated GO nano flakes as a tunable anisotropic liquid crystal.
The measurement method is based on the critical TIR angle of GO flowing in a micro-channel. In order to achieve this purpose, we make an experimental setup which is schematically shown in Figure \ref{fig:fig1}. The TEM00 mode of an intensity stabilized 10 mW Nd:YAG laser at 532 nm is passed through the polarizer, whose pass axis can be adapted to be TE or TM mode corresponding to s or p polarization, respectively. A half-wave ($\lambda/2$) plate is also used to adjust the optical power in the polarization direction of TM or TE mode. The laser beam is expanded and collimated by a collimator and then it is illuminated on a BK7 prism (with a refractive index of N=1.5) and we place a micro-channel on one of its faces as indicated in Figure \ref{fig:fig1}.

To explore the effect of flow rate on the birefringence properties of GO-LC,
 we utilize the micro-channel. Accordingly, we are able to control the flow of the liquid and the direction of the GO-LC director. We consider a LC between two parallel plates which is a homogeneous uniaxial anisotropic medium in which the optic axis is in the direction of the flakes (see Figure \ref{fig:fig1}(b) to make more sense). In this case, if a collimated beam is illuminated on the interface between two transparent semi-infinite media (one of these media is GO-LC with ordinary and extraordinary refractive indices of $n_o$ and $n_e$, respectively, and the other is an isotropic environment with a refractive index of N), therefore some part of mentioned beam are partially reflected and transmitted. If the optical axis is in the plane of incidence, the reflectivity can be obtained for the parallel polarizations (TM mode) and for the perpendicular (TE mode), based on the Maxwell equations as follows \cite{riviere1978}: 

\begin{eqnarray}
\label{eqn1}
R_{\parallel}=\left[ \frac{N\sqrt{ n_o^2 \cos^2 \alpha + n_e^2 \sin^2 \alpha - N^2\sin^2\delta}-n_e n_o \cos\delta}{N\sqrt{ n_o^2 \cos^2 \alpha + n_e^2 \sin^2 \alpha - N^2\sin^2\delta }+n_e n_o \cos\delta}\right]^2
\end{eqnarray}
\begin{eqnarray}
\label{eqn2}
R_{\perp}&=& \left[ \frac{N\cos\delta-\sqrt{ n_o^2-N^2\sin^2\delta}}{N\cos\delta+\sqrt{ n_o^2-N^2\sin^2\delta}}   \right]^2
\end{eqnarray}

where $\alpha$ is the angle between the optical axis and interface plane and $\delta$ is the incidence angle of the illuminated light. It is mentioned that the critical angle does not depend on the orientation of the optic axis $\alpha$ for perpendicular polarization $(\sin\delta_0=n_0/N)$, while, for parallel polarization, the critical angle depends on $\alpha$ and it is given by \cite{riviere1978}:

\begin{eqnarray}
\label{eqn3}
\delta_e(\alpha)=\sin^{-1}\left(\frac{1}{N}\sqrt{ n_o^2 \cos^2 \alpha + n_e^2 \sin^2 \alpha}\right)
\end{eqnarray}
We intend to use these relations to calculate the refraction indices of a GO-LC empirically. By injection of the GO-LC into the micro-channel, almost discotic flakes of GO-LC will be parallel to the micro-channel walls due to the anchoring force and cooperative alignment between LC flakes because of the aspect ratio of channel (Figure \ref{fig:fig1}(b)). So, it is possible to approximate the directors in a direction perpendicular to the interface between the two media, as shown schematically in Figure \ref{fig:fig1}(b). Therefore, the angle $\alpha$ can be assumed as $\alpha=90^\circ$, and we can write the relation for ordinary and extraordinary RI as follows:
\begin{eqnarray}
\label{eqn4}
\nonumber
\sin\delta_o=\frac{n_o}{N}  \\
\sin\delta_e(90^\circ)=\frac{n_e}{N}
\end{eqnarray}

\begin{figure}[ht]
\centering
\includegraphics[width=0.32\textwidth]{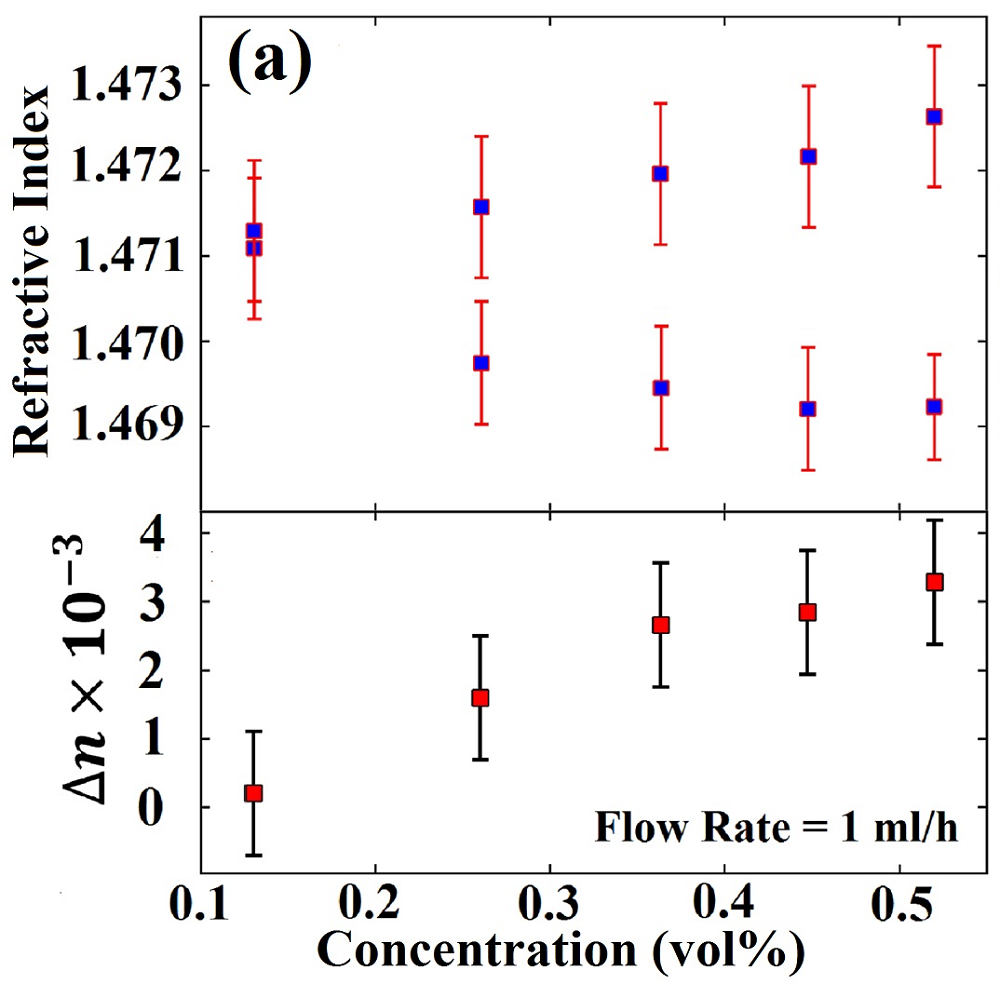}\\
\includegraphics[width=0.32\textwidth]{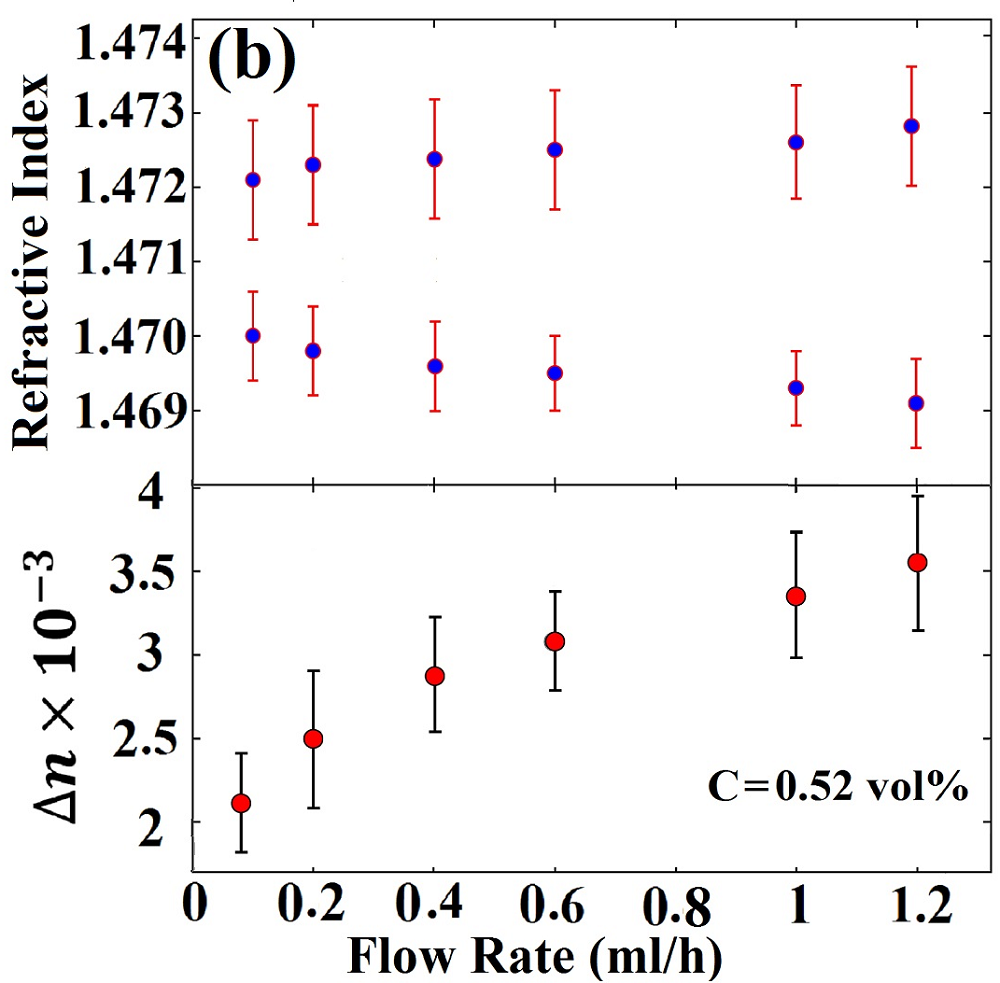}\\
\includegraphics[width=0.35\textwidth]{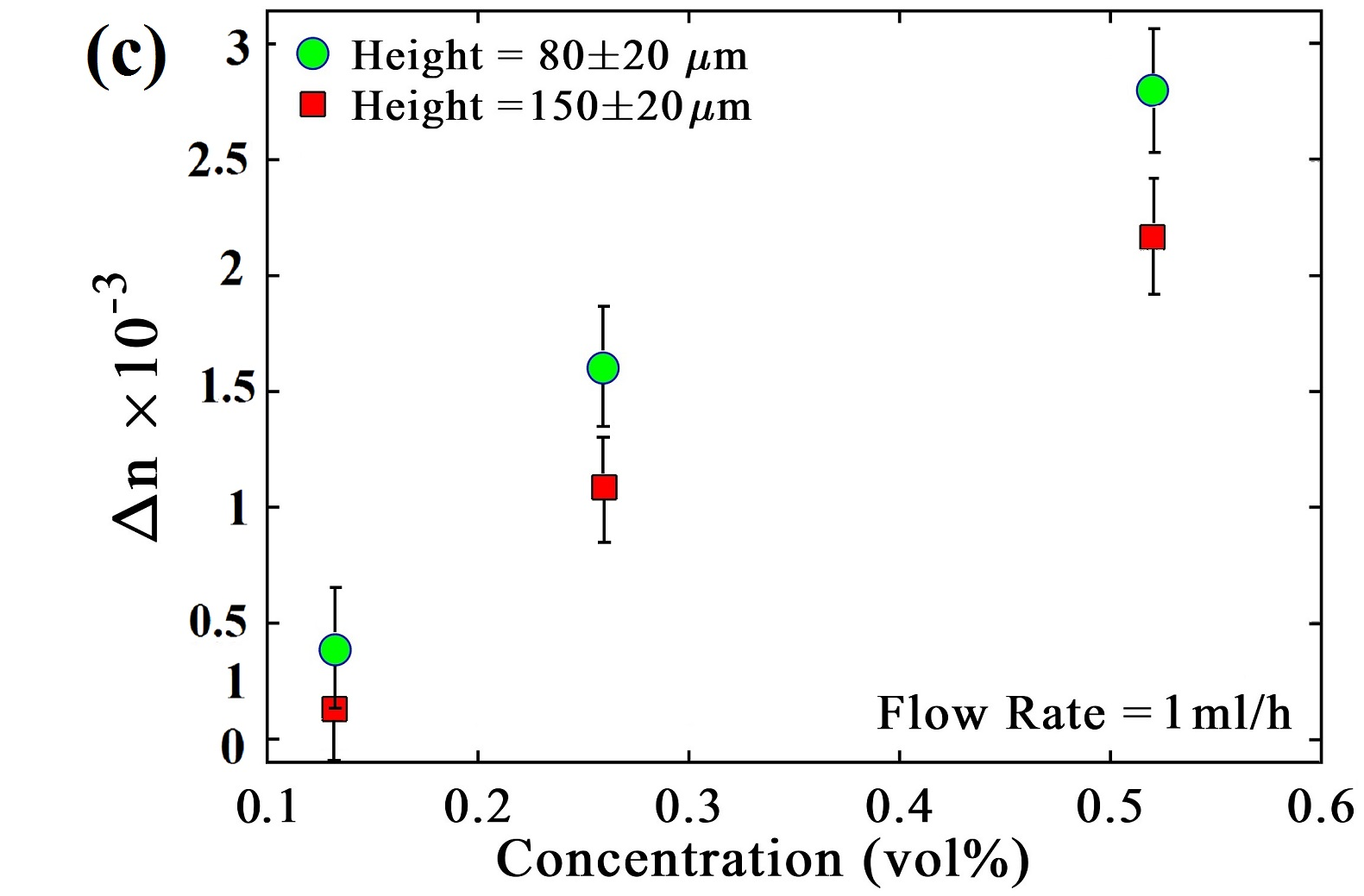}
     \caption{(a) Upper panel corresponds to the ordinary and extraordinary RI coefficients in terms of concentration. Lower panel indicates the controllable flow-induced birefringence for various concentrations of GO at flow rate of 1 $ml/h$, (b) Upper panel represents the RI coefficients values, $n_o$ and $n_e$, for the different flow rates. The lower panel illustrates the controllable flow-induced birefringence for various flow rate of GO at concentration of 0.52 $vol\%$, (c) flow-induced birefringence of GO for two different aspect ratio at flow rate of 1 $ml/h$ versus various concentration.}
\label{fig:fig2}
\end{figure}
We measured the critical angle for both polarization of s and p for five different concentrations of GO-LC at flow rate of 1 $ml/h$ and obtained $n_o$ and $n_e$ according to Eqs. \ref{eqn4}. The upper panel of Figure \ref{fig:fig2}a illustrates the values of $n_o$ and $n_e$ for the five different concentrations of GO-LC at the applied flow rate of 1 $ml/h$, while lower panel corresponds to $\Delta n\equiv n_e-n_o$ as a function of concentrations of GO-LC at the same applied flow rate. The higher value of GO-LC concentration leads to increase the $\Delta n$ which is associated with the anisotropy of LC. For this observation, when the GO concentration is sufficiently low, isolated flakes are the dominant form existing in the fluid, and flake-to-liquid is the dominant interactions. By increasing the concentration, the impact of the flake-to-flake interactions has increased which induce the GO flakes start to orient \cite{arshadi2019}. Increasing the alignment of flakes rise the order parameter of liquid crystals that leads to a growth the media birefringence \cite{bedford1994}.\\
For the further assessment, we examine the birefringence value, $\Delta n$, in terms of fluid flow rate. To this end, we inject GO-LC with different flow rates into the microchannel and determine the ordinary and extraordinary RI. The upper panel of Figure \ref{fig:fig2}b shows the ordinary and extraordinary RI as a function of flow rate for concentration of 0.52 $vol\%$, while lower panel correspond to birefringence for GO-LC at different flow rates. By increasing the flow rate, the $\Delta n$ represents a growing behavior. Indeed, by increasing the flow rate, the alignment of flakes direction is increased which leads to achieve more anisotropic behavior in media. It is worth noting that, applying external electric field is able to change the birefringence of the GO-LC as reported in \cite{shen2014,ahmad2015,ahmad2016}. But the amount of variation in the birefringence is less than that case when we use mechanical-hydrodynamical approach. However, one cannot compare our results with that of noticed in \cite{shen2014,ahmad2015,ahmad2016}, due to difference in experimental setup and material. Utilizing the induced external electric field causes to an undesired accumulation of GO at the electrode, while in mechanical-hydrodynamical, mentioned disadvantage would be irrelevant.\\

To investigate the effect of channel aspect ratio on birefringence, we consider two channels at thicknesses of $80\pm20 \mu m$ and $150\pm20 \mu m$, and for each channel, we perform the experiment at flow rate of 1 $ml/h$ for three different concentrations. The results are shown in Figure \ref{fig:fig2}c which demonstrates that by increasing the thickness of the channel (decreasing aspect ratio), the birefringence decreases, due to inducing the fluctuation of graphene oxide flakes in the far from boundaries region. On the other hand, the aspect ratio of the channel is another constraint that leads to the regular orientation of the GO flakes. 
\begin{figure}[ht]
\centering
\includegraphics[width=0.33\textwidth]{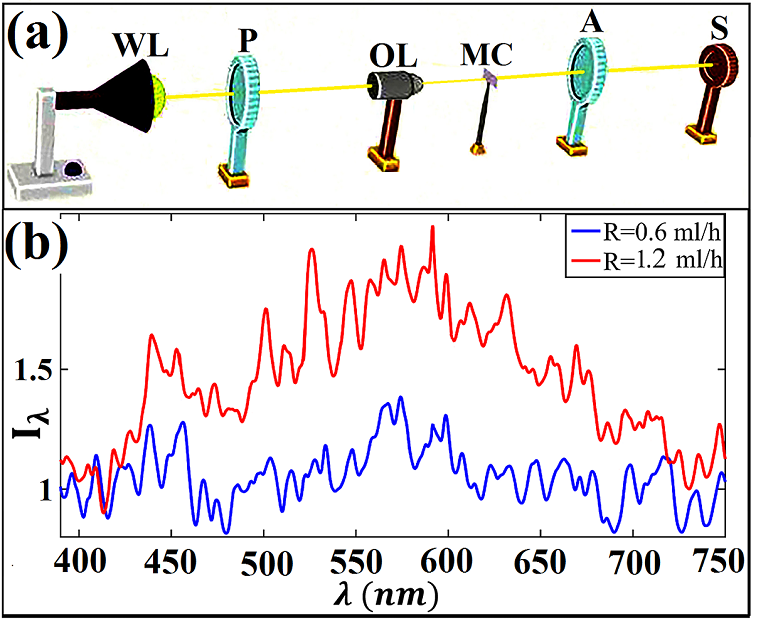}\\
\includegraphics[width=0.335\textwidth]{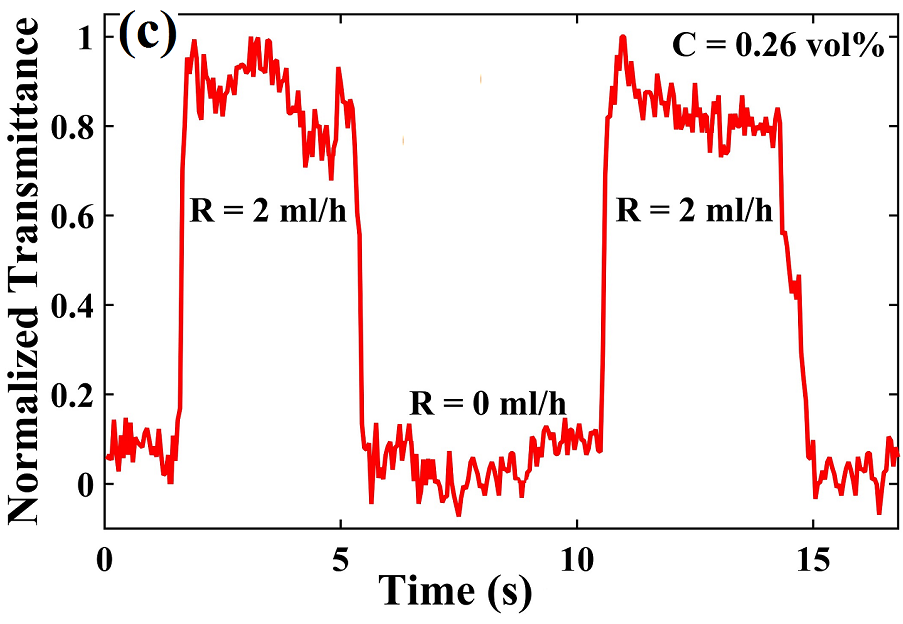}
     \caption{(a) Schematic of the experimental setup, arrangement of white light (WL) with a crossed polarizer. P, Polaroid; L, objective lens; MC, microfluidic channel; A, analyzer; D, detector (Spectrometer). (b) The output intensity of the transmitted light through a microfluidic channel containing GO-NLC for injection rates of 0.6 and 1.2 $ml/h$, where the channel is placed between crossed polarizer, (c) Normalized transmittance laser beam through levitated GO's flakes at concentration of 0.26 $vol\%$ in the microchannel with and without existence of flow rate (2 $ml/h$) as an optical switching.}
\label{fig:fig3}
\end{figure}

Focusing the polarized white light on the sample and examining the transmitted rays passing from an analyzer as illustrated in Figure \ref{fig:fig3}a can recognize the effect of injection rate on the birefringence of GO-LC. To prove this idea, we use a white light lamp with the spectral region from 400-800$nm$. To detect the light passing through the analyzer, we utilize a spectrometer. Figure \ref{fig:fig3}b shows the measured normalized transmitted intensity ($I_\lambda=(I_R-I_0)/I_0$, $I_R$ and $I_0$ are the intensity of light with and without injection rate, respectively) collected after the analyzer for two injection rates of 0.6 and 1.2 $ml/h$. By increasing the injection rate of LC into the channel, the intensity of the beam passing through the analyzer grows. In other words, the GO-LC sample changes the linear polarization of the light into elliptically polarization and by increasing the birefringence due to the flow rate, the variation of the elliptically polarization states become large. As a result, the intensity of the transmitted light from the analyzer is changed at different wavelength. \\
The proposed method can be used as an optical switch. For this purpose, we use the setup of Figure \ref{fig:fig3}a. We only use laser instead of white light. The result of this experiment is shown in Figure \ref{fig:fig3}c. It is observed that applying the flow rate acts as a switch and causes the laser light to pass through the analyzer. By improving the orientation of the GO's flakes and the functionalization of the channel wall surfaces, the switching feature of the microfluidic systems containing GO-LC can be improved. This effect can be used to design a mechanical-hydrodynamical switch in which the intensity of the output light controlled by manipulating the flow rate. This mechanical-hydrodynamical system can be used for cases in which the electric and magnetic fields cannot be used to control the birefringence.\\
In summary, in this work, we have successfully measured the values of both the ordinary and extraordinary refractive indices of the graphene oxide liquid crystals. Examining the transmitted polarized light across the GO-LC samples revealed a relation between the birefringence and concentration. We have shown that by changing the concentration and flow rate of the GO-LC sample in the micro-channel, the value of ordinary and extraordinary as well as the birefringence of the samples changed. We found out that the overall behavior for dependency of birefringence variations to concentration and flow rate when we use white light is similar to that of observed with monochromatic incident light. Our results reveal the occurrence of a controllable birefringence GO nano flakes material forward promising LC to be applied in the optical switch.

\begin{acknowledgements}
The authors would like to thank M. M. Jahan-bakhshian and H. Sepahvand for their comments on the manuscript.

\end{acknowledgements}


\end{document}